# Quadratic heat capacity and high-field magnetic phases of $V_5S_8$

C. A. Sonego[1,2], P. M. T. Vianez[1], H. Li[1], J. C. Lashley[1], G. G. Lonzarich[1], M. A. Avila[2] and S. E. Rowley[1,*]

1. Cavendish Laboratory, University of Cambridge, J. J. Thomson Avenue, Cambridge, CB3 0HE, United Kingdom
2. CCNH, Universidade Federal do ABC, Santo André, São Paulo, Brazil
* The corresponding author's email address is ser41@cam.ac.uk

**We report the observation of an unexpected quadratic temperature dependence of the heat capacity in the vanadium sulphide metal $V_5S_8$ at low temperatures which is independent of applied magnetic field. We find that the behaviour of the heat capacity is consistent with an unconventional phonon spectrum which is linear in wavevector in the *c* direction but quadratic in the *a-b* plane, indicating a form of geometrical elastic criticality. In the case of $V_5S_8$ we also observe an unusual intermediate transition at high magnetic fields between the expected spin-flop and spin-flip transitions. We demonstrate that the intermediate field-induced transition is in agreement with a model of two sublattices with frustrated inter- and intra-sublattice spin couplings.**

## 1. Introduction

The chalcogenides $V_xS_8$, in which *x* can vary between 4 and 8, form an interesting family of narrow-band d-electron metals that host a wide range of charge and magnetic order at low temperatures upon variations of stoichiometry and pressure. $VS_2$ (*x*=4) is a charge-density-wave metal below approximately 300K [1,2], $V_5S_8$ (*x*=5) and $V_3S_4$ (*x*=6) are antiferromagnetic metals with Néel temperatures of approximately 32K and 8K respectively, and VS (*x*=8) is a paramagnetic metal [3–9] (see Fig. 1). The family may be expanded further with Se or Te in place of S and other transition-metal elements in place of V. $VS_2$ is of interest due to its close relation to $VO_2$, a strongly correlated electron system that becomes a Mott Insulator below 340K [10]. Under suitable conditions the itinerant carriers in vanadium sulphides may condense into unconventional forms of superconductivity, including those anticipated to emerge on the border of magnetism [11] and charge-density-waves [12,13]. The vanadium sulphides have also been identified as an important platform for energy technologies due to their excellent electrochemical properties and subsequent performance as electrode materials in metal-ion batteries and supercapacitors [14,15].

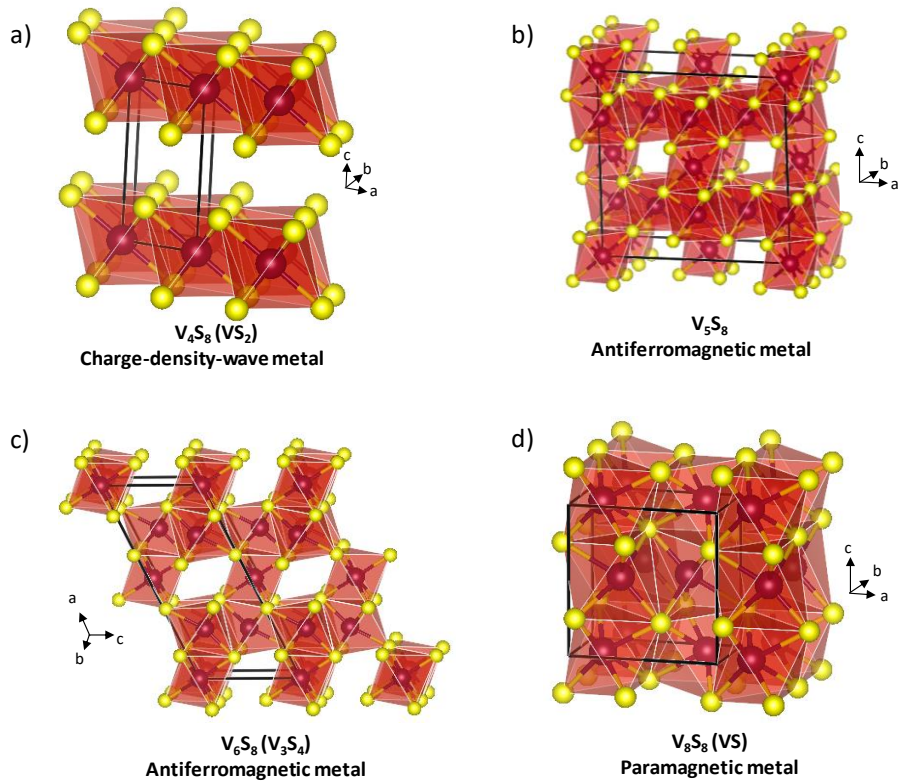

**Figure 1** – The crystal structures of the $V_xS_8$ family of d-electron metals for *x* running through values 4, 5, 6 and 8. The vanadium ions are depicted as purple spheres and the sulphur ions as yellow spheres. The black lines indicate the unit cells of the different crystal structures shown and the orientations depicted are denoted by the crystallographic axes labels *a*, *b* and *c*. The faces of polyhedra formed by sulphur ions which enclose individual vanadium ions are shaded in red. The crystal structure space groups of the different family members are in (a) P$\bar{1}$m1, trigonal ($VS_2$), in (b) F 1 2/m1, monoclinic ($V_5S_8$), in (c) C1 2/m 1, monoclinic ($V_3S_4$) and in (d) Pcmn, orthorhombic (VS). $VS_2$ is a layered material that may be thought of as having a quasi two-dimensional aspect and for which additional vanadium may be intercalated inside the van der Waal gaps forming the other materials shown.

$VS_2$ may be prepared as a two-dimensional material when grown as thin molecular layers by molecular beam epitaxy (MBE) [1,14]. The $VS_2$ monolayers form a honeycomb structure reminiscent of graphene but with one of the two inequivalent sites occupied by V and the other essentially by $S_2$ (Figs. 2a & 2b). In bulk form, the $VS_2$ layers are held together by weak van der Waal forces (Fig. 1a) and atoms such as V, Ag, Ti, Mo, Nb and Ta may be intercalated between the layers with different degrees of mobility. Indeed, $V_5S_8$ and $V_3S_4$ may be thought of as particularly stable and ordered phases of $VS_2$ intercalated with additional vanadium between the layers. This results in metal-full layers and metal-deficient layers as shown in Fig. 1. Antiferromagnetic $V_5S_8$ and $V_3S_4$ have been of long standing interest in the field of metallic magnetism where the degree of itinerancy of the magnetic electrons along with the importance of spin-fluctuation corrections to the thermal properties have been investigated [4,5,22,23,6,9,16–21]. In the case of $V_5S_8$, nuclear magnetic resonance [7] and neutron scattering [8] experiments have identified the presence of three different types of vanadium ions, $V_I$ between the layers and $V_{II}$ and $V_{III}$ within the layers. The $V_I$ sites alone predominantly host magnetic electrons which order below the Néel temperature, $T_N$, with an orthorhombic (slightly distorted face-centred-tetragonal) spin lattice as shown in Fig. 2c. The low temperature atomic moment is of the order of $1.5\mu_B$ per $V_I$ site so that a description in terms of a Heisenberg spin model may be appropriate to a first approximation for the bulk properties. However, as in the case of magnetic metals such as iron and cobalt with equally large moments, the Fermi surface is likely to be described by an itinerant-electron or Stoner model [17,21–23].

As a first step in the search for exotic condensates in the magnetic vanadium chalcogenides we have carried out measurements of principal thermal properties of stabilized samples of $V_5S_8$ at ambient pressure. The measurements summarized below reveal an unexpected and anomalous temperature dependence of the heat capacity $C(T)$. In particular, the ratio $C/T$ is not quadratic in $T$ as is often observed but approximately linear from mK temperatures up to around 15K and essentially independent of magnetic field up to at least 14T. This quadratic temperature dependence, $C \sim T^2$, in the low temperature limit is a very rare result in the study of solid-state single-crystalline materials. We see later that these results appear to be consistent with a form of geometrical elastic criticality usually only present in monolayer honeycomb structures that give rise to unconventional acoustic phonon spectra. Furthermore, instead of the two field-induced magnetic transitions predicted by the spin-flop and spin-flip model, in $V_5S_8$ we observe three transitions for magnetic fields aligned along the easy axis of magnetization, $c$, below $T_N$. The possible origins of these unexpected findings will be considered and discussed below after presenting the results of magnetic, electrical transport and thermal measurements.

## 2. Magnetisation, electrical transport and heat capacity results

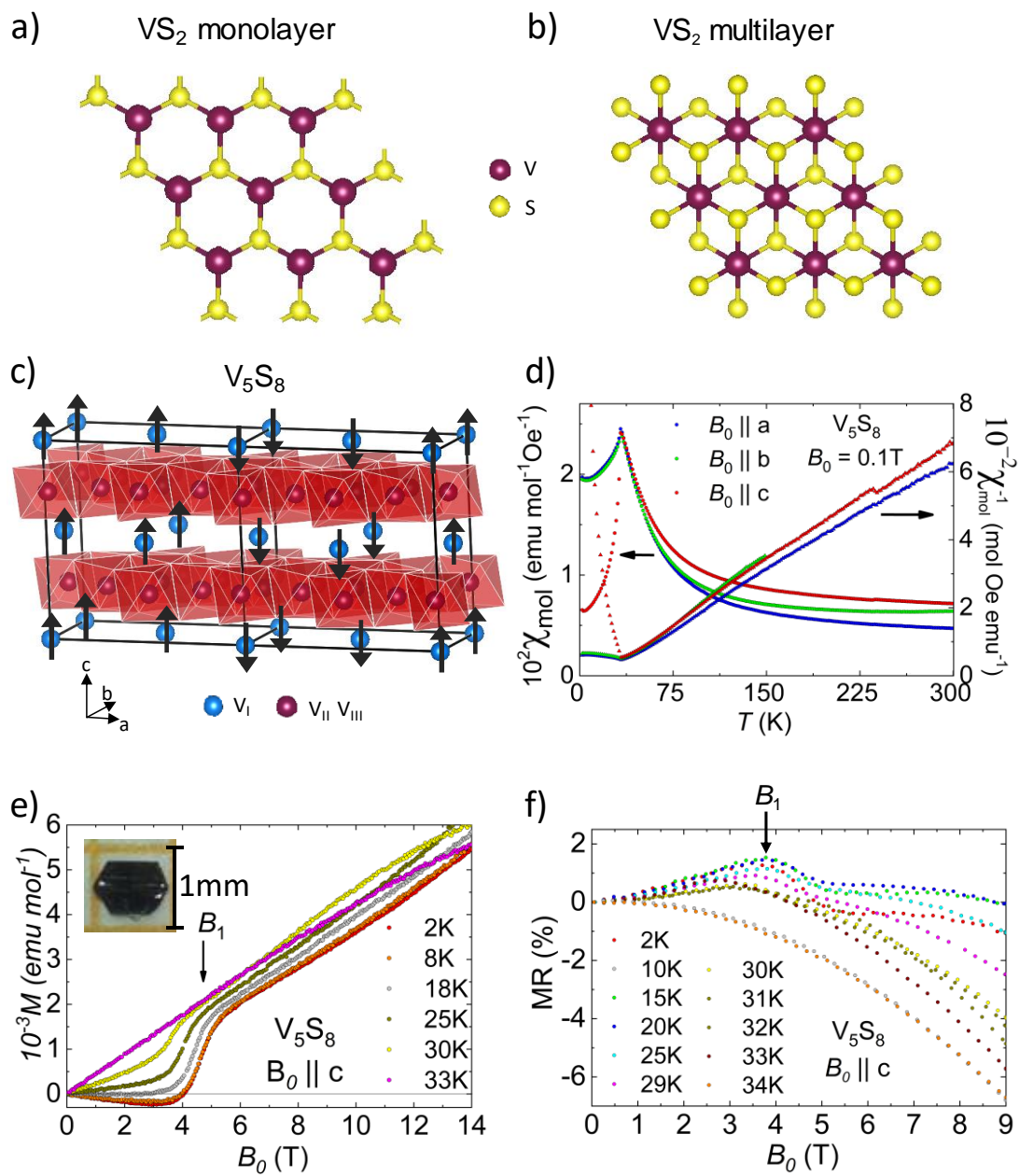

**Figure 2** – In (a) the H-VS$_2$ crystal structure is shown for a monolayer of VS$_2$ and in (b) for the T-VS$_2$ structure of a layer in bulk VS$_2$ [14], i.e. as viewed from the top along the *c*-direction in the image shown in Fig. 1a. Additional vanadium may be intercalated between the layers of bulk VS$_2$ resulting in V$_5$S$_8$ and V$_3$S$_4$ as shown in Figs. 1b & 1c which comprise metal-full and metal-deficient layers. The magnetic structure of V$_5$S$_8$ below the Néel temperature [8] is shown in (c) in which magnetic electrons represented by arrows reside on the V$_I$ sites found in the metal-deficient layers. (The sulphur ions and polyhedra surrounding V$_I$ sites have been hidden for clarity.) In (d) the magnetic susceptibility, $\chi_{mol}$, of V$_5$S$_8$ versus temperature is plotted for the three principal crystallographic directions at a low applied magnetic field. In (e) the magnetization, *M*, and in (f) the magnetoresistance, $MR = (R(B_0,T) - R(0,T)) / R(0,T)$, where *R* is the resistance, of V$_5$S$_8$ is plotted against applied magnetic field aligned along the easy, *c*, axis at a series of temperatures up to just above $T_N \approx 32$K. The spin-flop transition field, $B_1$, is visible

in both of these quantities. The quantities in (d) & (e) are plotted per mole of $V_5S_8$ which has a relative molecular mass, $M_r$ = 511.23 g/mol and density, $\rho$ = 3.95g/cm$^3$. The conventional cell contains four $V_I$ ions out of a total of twenty vanadium ions, and one $V_I$ ion per $V_5S_8$ molecule in the primitive cell.

Crystals of vanadium sulphide metals were prepared by the vapour transport technique [24–26] and were confirmed by X-ray crystallography experiments (including X-ray diffraction (XRD), and energy-dispersive X-ray spectroscopy (EDS)) to form in a single phase with the conventionally accepted lattice structure and stoichiometry (see also Methods and Appendix A). The magnetic susceptibility as a function of temperature, $T$, and of the magnetization as a function of applied magnetic field, $B_0$, along the three principal crystallographic axes were measured using a commercial superconducting quantum interference device magnetometer. The results for $V_5S_8$ are shown in Figs. 2d, 2e & 2f. The susceptibility plotted against $T$ in Fig. 2d shows moderate anisotropy above $T_N$ and strong anisotropy below $T_N$ where the susceptibility along the easy, $c$, axis falls below that along the $a$ and $b$ axes, as expected for an antiferromagnetically ordered state. The magnetization versus $B_0$ along the $c$-axis plotted in Fig. 2e at a series of temperatures shows a steep rise at a critical field of about 4.5 T in the low temperature limit ($T \ll T_N$), qualitatively as expected for spin-flop transitions which have been observed in a number of antiferromagnetic materials. Signatures of this transition have also been observed in transport properties as shown in Fig. 2f. The temperature and field dependence of the heat capacity, however, is suggestive of additional field-induced transitions.

The heat capacity of $V_5S_8$ as a function of $T$ and $B_0$ were measured using a standard relaxational technique and the results are presented in Fig. 3. The heat capacity plotted against $T$ in Fig. 3a shows a sharp peak which falls in amplitude and temperature as $B_0$ is increased. In the low field limit the peak appears at approximately at 32K, the accepted value of the Néel temperature. As shown on an expanded scale in Fig. 3b, however, the heat capacity peak splits apart with increasing $B_0$ pointing to the existence of an unexpected additional transition line that can also be detected in the resistivity as in Fig. 3c. The magnetic field–temperature phase diagram shown in Fig. 3d, which is compiled from features in the magnetization, heat capacity and resistivity data, shows the presence of three transition lines labelled $B_1$, $B_2$ and $B_3$ that decrease with increasing temperature and all meet together near $T_N$. The triad of transition lines is not a widely observed characteristic of antiferromagnetic materials.

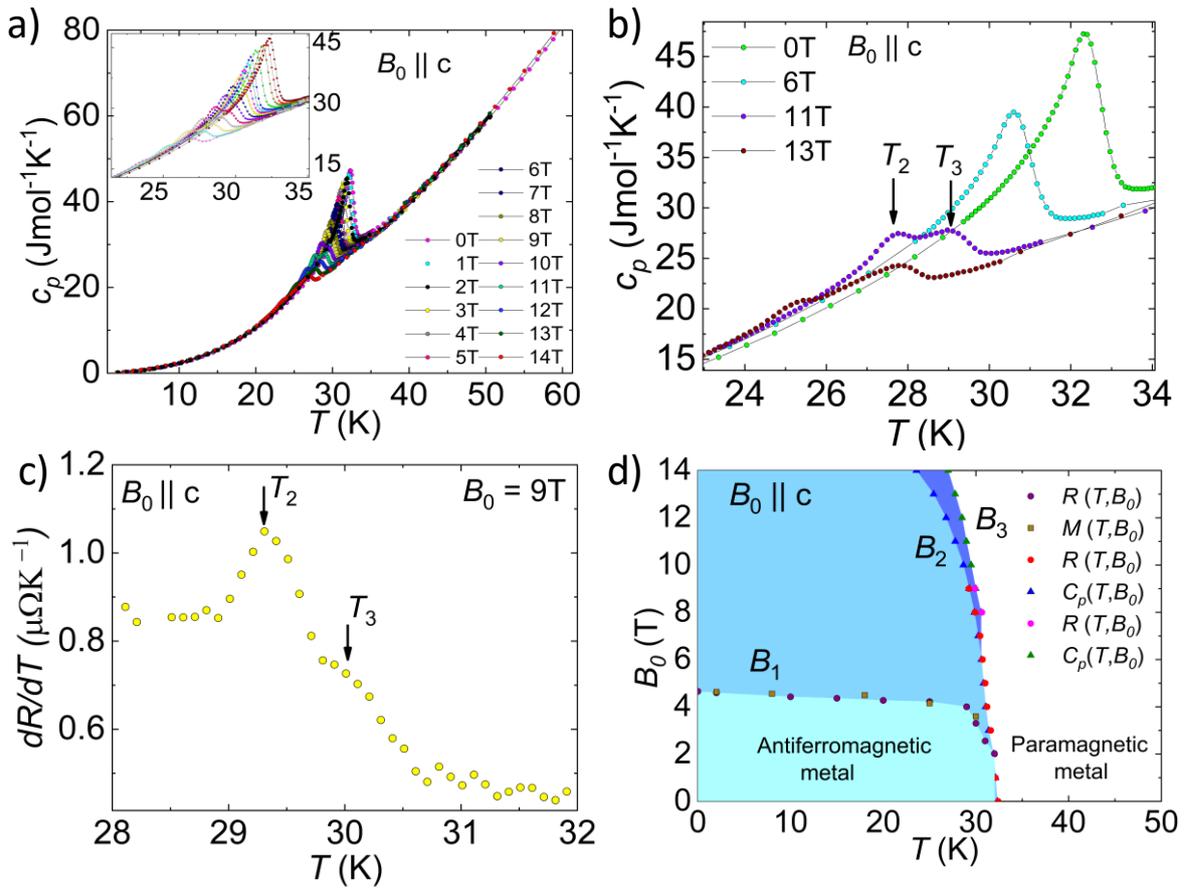

**Figure 3 – Heat capacity, resistance and phase diagram of $V_5S_8$.** In (a) the heat capacity vs. temperature is plotted at a series of applied magnetic fields aligned along the easy, $c$, axis. The inset is a plot of the heat capacity data in the vicinity of $T_N$. In (b) the heat capacity vs. temperature is plotted at a series of applied magnetic fields aligned along the easy, $c$, axis in the vicinity of $T_N$ highlighting the splitting of the transition in applied magnetic fields as shown by peaks in temperature labelled $T_2$ and $T_3$. The same peaks in $T$ are also observed in the derivative of the resistance with respect to temperature ($dR/dT$) as shown in (c). In (d) the applied magnetic field vs. temperature phase diagram showing the temperature dependence of the three field induced transitions, $B_1$, $B_2$ and $B_3$, associated with peaks $T_1$, $T_2$ and $T_3$ observed in the measured quantities labelled. ($B_1$ is obtained from the resistance and magnetisation data as presented in Fig. 2.) The quantities in (a) & (b) are plotted per mole of $V_5S_8$ which has a relative molecular mass and density as noted in the caption to Fig. 2.

Even more unusual is the quadratic, $T^2$, variation of the heat capacity at low temperature for $V_5S_8$ as plotted in Fig. 4a, which is weakly dependent on $B_0$ at least up to 14T, i.e., well above the magnetic transition field line $B_1$, the spin-flop transition. The weak field dependence suggests that the heat capacity is mainly due to lattice vibrations that are normally expected to give rise to a cubic, $T^3$, contribution to the heat capacity far below the Debye temperature.

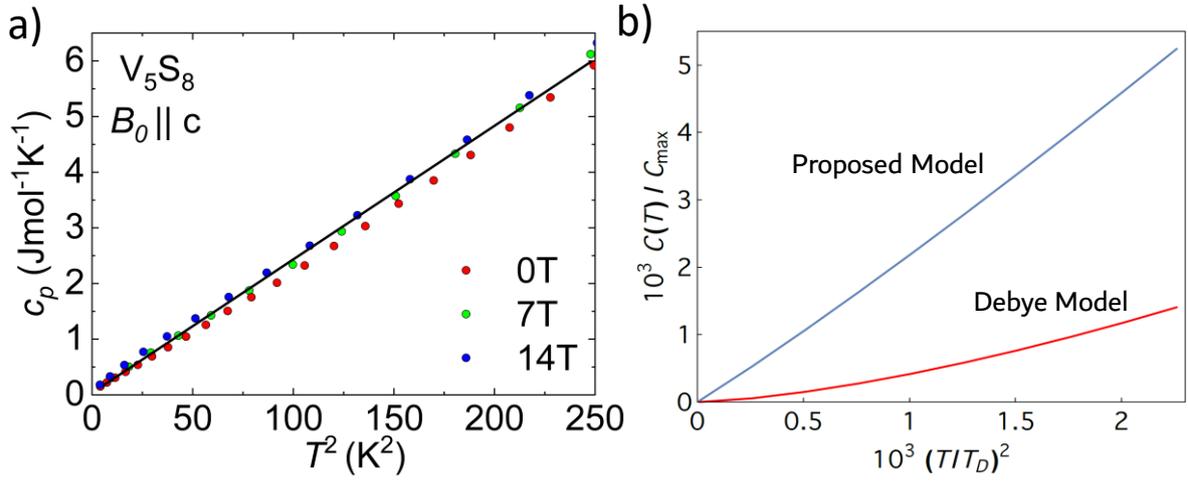

**Figure 4** – Quadratic temperature dependence of the low temperature heat capacity. In (a) the measured heat capacity vs. $T^2$ is plotted at low temperatures for $V_5S_8$ for a series of applied magnetic fields aligned along the easy, $c$, axis. The measured value is plotted per mole of $V_5S_8$ which has a relative molecular mass and density as noted in the caption to Fig. 2. The heat capacity plotted against $T^2$ shown in (b) is the result of a model calculation based on acoustic phonon energy spectra of the form $\epsilon(q) = q_x^\alpha + q_y^\beta + q_z^\gamma$, where $\epsilon$ and $q$ are in units of the Debye energy and Debye wavevector respectively. In the proposed model discussed in more detail in the main text $\alpha = \beta = \gamma = 1$ for two acoustic modes, and $\alpha = \beta = 2$ and $\gamma = 1$ for the third mode. In the conventional Debye model $\alpha = \beta = \gamma = 1$ for all three acoustic modes. A more realistic model based on a series expansion of the square of the phonon energy to second and fourth order in $q_x, q_y$, and $q_z$ leads to similar results for the qualitative form of the anisotropy postulated above.

## 3. Quadratic temperature dependence of the heat capacity

The weak field dependence of the $T^2$ heat capacity even in fields far above the magnetic transition field $B_1$ suggests that its origin is not mainly due to spin fluctuations but rather of lattice vibrations. We consider the effects of deviations from the usual linear dispersion of the acoustic phonon spectrum. If the square of the phonon energy is an analytic function we would expand it in a series in the cartesian components $q_x$, $q_y$, $q_z$ of the wavevector consistent with the symmetry of the lattice. For our purposes it is sufficient to consider instead a simple dispersion relation of the form $\epsilon(q) = aq_x^\alpha + bq_y^\beta + cq_z^\gamma$ where $a, b, c$ and $\alpha, \beta, \gamma$ are phenomenological parameters. The thermal population given by the $q$-space integral of the Planck function, $n_T = (1/8\pi^3) \int dq/(\exp(\epsilon(q)/T) - 1)$, yields the qualitative temperature dependence of the heat capacity at low $T$.

In leading order in $T$ the heat capacity exponent is thus expected to be $\nu_{3D} = (\frac{1}{\alpha} + \frac{1}{\beta} + \frac{1}{\gamma})$ for a given acoustic mode. For a single layer normal to the $z$ axis ($c$ axis) we drop the $z$ dependence in the integral and the exponent reduces to $\nu_{2D} = (\frac{1}{\alpha} + \frac{1}{\beta})$. For a monolayer of the parent material VS$_2$, the longitudinal acoustic mode and the transverse acoustic mode for displacements of the atoms within the layer lead to a linear spectrum ($\alpha = \beta = 1$) so that the exponent $\nu_{2D} = 2$. However, for the transverse acoustic mode due to displacements of the atoms normal to the layer, the spectrum is quadratic ($\alpha = \beta = 2$) so that the exponent $\nu_{2D} = 1$. The latter flexural modes (also known as ZA phonons) produce the dominant contributions at low $T$ so that the heat capacity is expected to be linear in $T$. The quadratic spectrum for transverse-acoustic phonons involving motions of the atoms normal to the plane is a special feature of the geometry of the honeycomb crystal structure of VS$_2$ [27–29] (Fig. 2a) and related materials such as graphene [30–33]. The effective spring constants of neighbouring springs that model the lattice vibrational modes are found to balance out in these special geometries resulting in soft acoustic phonons.

In a multilayer structure the coupling along the $z$ axis is expected to lead to linear dispersions for all three acoustic modes in all three directions ($\alpha = \beta = \gamma = 1$) so that the exponent $\nu_{3D} = 3$. However if the flexural mode of the monolayer transforms in the weakly coupled multilayer system into an acoustic mode with quadratic dispersion in the $x$ and $y$ directions but linear dispersion along the $z$ direction [34] ($\alpha = \beta = 2$, $\gamma = 1$) then $\nu_{3D} = 2$, and the dominant term in the heat capacity at low $T$ would be expected to be quadratic in $T$, as observed (Fig. 4a). The heat capacity can be calculated based on acoustic phonon spectra of this form, i.e. $\epsilon(\mathbf{q}) = q_x^\alpha + q_y^\beta + q_z^\gamma$ where $\epsilon$ and $q$ are in units of the Debye energy and Debye wavevector respectively. In the proposed model involving flexural modes, $\alpha = \beta = \gamma = 1$ for two acoustic modes, and $\alpha = \beta = 2$ and $\gamma = 1$ for the third mode. The resulting heat capacity is plotted against $T^2$ in Fig. 4b. A plot of the heat capacity based on the conventional Debye model is also shown in which $\alpha = \beta = \gamma = 1$ for all three acoustic modes. The model outlined here therefore not only accounts for the temperature dependence of the heat capacity but also the enhanced absolute value of the heat capacity beyond that expected from the Debye model. This is one possible origin of the quadratic heat capacity observed in V$_5$S$_8$. We also note that numerical calculations of the phonon density of states of lattice vibrations in the hypothetical fully randomized tungsten-rhenium system, have revealed that over a range in composition where the lattice structure is metastable, the density of states is linear in energy [35]. The calculated phonon spectrum for W$_{1-x}$Re$_x$ near $x = 0.5$ is linear along [100], but appears roughly quadratic along [110] and [111]. The expected heat capacity in this model is also quadratic in $T$. Quadratic phonon dispersions occurring in particular crystal directions can therefore arise in materials beyond those supported by monolayer honeycomb structures.

In both V$_5$S$_8$ and W$_{1-x}$Re$_x$ this unusual spectrum may be connected with the presence of frustration in the nearest-neighbour and next-nearest-neighbour couplings resulting in cancelling of effective spring constants. In certain cases the effect of positional ion disorder may also be relevant. A quadratic acoustic spectrum is usually only found in materials tuned to the critical point of certain elastic phase transitions [36,37] but appears here to be preserved over wide ranges in temperature due to the geometry of the crystal structure. The proposal for the origin of the quadratic heat capacity and elastic criticality in vanadium sulphides may be tested in particular by inelastic neutron scattering measurements of the lattice excitations. It may also be feasible to calculate the lattice vibrational spectrum in sufficient detail to look for deviations from the conventional linear dispersion of the acoustic modes at low wavevectors along different directions in reciprocal space.

## 4. The triad of transition lines in $V_5S_8$

We begin by considering a uniform antiferromagnetic chain extended along the $y$ direction with the easy axis of magnetization along the $z$ direction, in a mean field approximation. The ground state magnetic energy of a single antiferromagnetically aligned pair at neighbouring sites $i$ and $i-1$ in the presence of single-ion anisotropy and an applied magnetic field $B_0$ directed along the $z$ axis is taken to be of the form

$$\frac{E_{pair}}{g\mu_B<S>} = -B_{exc}cos(\theta_i - \theta_{i-1}) - \frac{1}{2}B_{an}(\cos^2\theta_i + \cos^2\theta_{i-1}) - B_0(cos\theta_i + \cos\theta_{i-1}) \quad (1)$$

where $<S>$ is the magnitude of the average spin on each site, $B_{exc} = -\frac{J<S>}{g\mu_B}$ is an exchange field that anti-aligns the two moments when the exchange coupling $J > 0$, and $B_{an}$ is the local uniaxial anisotropy field tending to align the spins along the $z$ axis [38,39]. The total ground state magnetic energy for the lattice $E_{lattice}$ in the mean field approximation is given by $E_{pair}$ times the number of pairs but with the exchange term doubled to take account of the exchange contribution from the bonds between the pairs. The variable $\theta_i$ is the angle between the $z$ axis and a spin at site $i$ which can rotate in the $z$-$y$ plane to minimise the total energy.

The spin pairs are anti-aligned for $B_0 \ll B_{exc}$ and aligned in the opposite limit but the transition is not continuous. Minimizing $E_{lattice}$ shows that for a range of parameters there exits a first order "spin-flop" transition to partial spin polarization at a value of $B_0$ that scales as $\sqrt{B_{an}B_{exc}}$ followed by a second order "spin-flip" transition to full spin polarization at $B_0$ that scales as $B_{exc}$ (Fig. 4a ). At the spin-flop transition the spin angles for a pair suddenly jump from $\theta_i = 0$, $\theta_{i-1} = -\pi$ to $\theta_i = \theta_{i-1} = \theta = \theta_{SF}$, where $0 < \theta_{SF} < \frac{\pi}{2}$. The angle $\theta$ continuously decreases with increasing $B_0$ from $\theta_{SF}$ but does not smoothly tend to zero at infinity; instead $\theta$ vanishes at a finite spin-flip transition field. At the spin-flop field, the applied field overcomes the tendency of the exchange field together with the anisotropy field to align the spins in a pair in opposite directions along the $z$ axis.

In the above model there are only two transitions. However further transitions can arise in more general forms of the pair energy. In particular, the inclusion of anisotropy in the two-spin and single-spin terms can naturally lead to three transitions, qualitatively as observed in $V_5S_8$. To see this we generalize $E_{pair}$ to the form

$$\frac{E_{pair}}{g\mu_B <S>} = -\left(B_z^{exc}\cos\theta_i \cos\theta_{i-1} + B_y^{exc}\sin\theta_i \sin\theta_{i-1}\right) - \frac{1}{2}B_z^{an}(\cos^2\theta_i +$$
$$\cos^2\theta_{i-1}) - \frac{1}{2}B_y^{an}(\sin^2\theta_i + \sin^2\theta_{i-1}) - B_0(\cos\theta_i + \cos\theta_{i-1}) \qquad (2)$$

where we have replaced $\cos(\theta_i - \theta_{i-1})$ by $\cos\theta_i \cos\theta_{i-1} + \sin\theta_i \sin\theta_{i-1}$ and included not only $z$ but also $y$ components of the two-site and single-site fields. Minimization of the generalized form of $E_{lattice}$ leads to two transitions if the single site anisotropy term tends to favour spin alignment along the $z$ axis ($B_z^{an} > B_y^{an}$), as shown in Figs. 5a & 5b. However, when the combined effects of the two-site and single-site terms make the $z$ axis the easy axis as above, but the single site term on its own favours spin alignment along the $y$ axis ($B_z^{an} < B_y^{an}$) then three phase transitions are predicted for a range of model parameters, as shown in Figs. 5c & 5d. In the latter case the conflicting tendencies of terms in $E_{lattice}$ for alignment along the $z$ axis and $y$ axis leads to a splitting of the first-order spin-flop transition into two second-order transitions $B_1$ and $B_2$. In the intermediate state between $B_1$ and $B_2$, the alignment of the two spins is given by $\theta_i = \zeta + \theta$ and $\theta_{i-1} = \zeta - \theta$, where $\zeta$ varies continuously from $\frac{\pi}{2}$ at $B_1$ to 0 at $B_2$. The spin-flip transition at still higher fields where $\theta$ also vanishes is labelled $B_3$.

The above chain model can be generalized to the case of two sublattices for which the two site energy is replaced by an inter-sublattice coupling and the single-site energy is replaced by an intra-sublattice energy, the phenomenological parameters generalized accordingly. The qualitative form of the resulting lattice magnetic energy in the mean field approximation is similar to that given above, leading to the same triad of magnetic transitions as shown in Figs. 5c & 5d in terms of phenomenological model parameters. This proposal for the possible origin of the three magnetic transitions provides only one direction to pursue and to go further requires a more realistic model of the rather complex and unusual spin structure of $V_5S_8$. Furthermore, this discussion provides some instruction for more penetrating experimental investigations in particular by means of elastic neutron scattering measurements of the magnetic structure of $V_5S_8$ as a function of applied magnetic field with polarized neutrons.

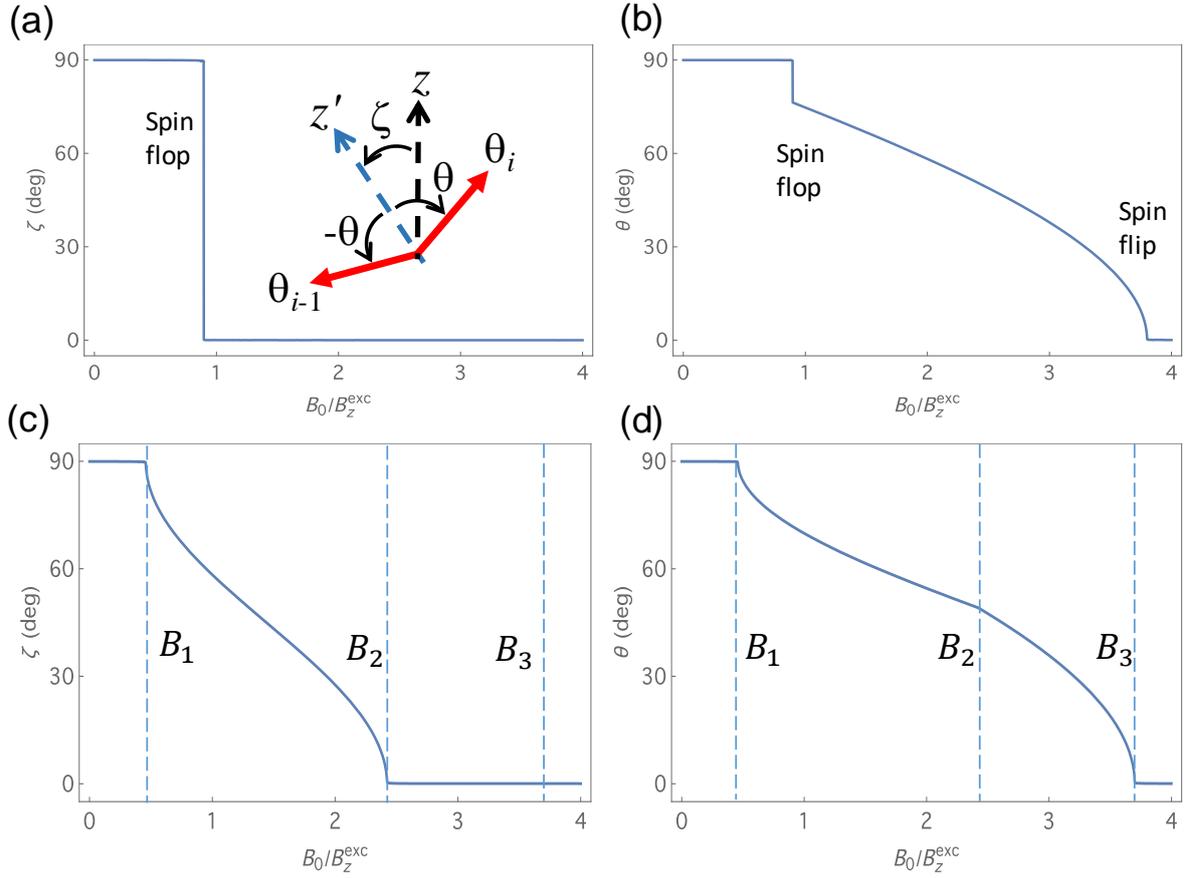

**Figure 5** – Magnetic field-induced transitions in the mean field approximation for the lattice spin model described in the main text including anisotropic inter-sublattice and intra-sublattice energies and the Zeeman energy. For illustration the model parameters have been chosen to be $B_y^{exc}/B_z^{exc} = 1.0$ and $(B_z^{an} - B_y^{an})/B_z^{exc} = 0.2$ in (a) and (b), and $B_y^{exc}/B_z^{exc} = 0.1$ and $(B_z^{an} - B_y^{an})/B_z^{exc} = -1.5$ in (c) and (d), and the results shown in the figures were obtained by numerical minimisation of the magnetic lattice energy $E_{lattice}$ for the ground state as outlined in the main text. The inset in (a) indicates the definitions of the angles θ and ζ. $θ_i$ is the angle of a spin at site $i$ as measured from the $z'$ direction. ζ is the angle of the $z'$ direction as measured from the $z$ axis. The orientations of the spins represented by red arrows are able to vary in the $z$-$y$ plane in order to minimise the energy. In the latter set of parameters in (c) and (d), the intra-sublattice coupling favours spin alignment along the $y$ axis and the first-order spin-flop transition splits into two second-order transitions leading to three field-induced second-order transitions $B_1$, $B_2$ and $B_3$ similar to those observed in V$_5$S$_8$ (Figs. 2 & 3).

## 5. Conclusions

The most striking findings are the quadratic temperature dependence of the low temperature heat capacity, which is nearly magnetic field independent. This appears to be a particularly rare experimental result in the study of bulk single crystals. The $T^2$ heat capacity points to the possible existence of an unconventional form of the energy spectrum of one or more of the acoustic phonon modes of V$_5$S$_8$ as outlined above that could be investigated by means of the inelastic scattering of neutrons from lattice vibrations. The proposed modes involve quadratic rather than linear transverse-acoustic phonon dispersions in particular crystallographic directions. Intriguingly an acoustic phonon energy spectrum of the form $\epsilon^2 = v_1^2 q^2 + v_2^4 q^4$ is appropriate for certain classes of ferroelastic materials in which the speed of sound vanishes, $v_1 \to 0$, in certain directions at the elastic critical point resulting in a quadratic dispersion [36,37]. From this point of view, the layered structures based on stackings of honeycomb $VS_2$ monolayers and related compounds can be thought of as hosting a form of geometrical elastic criticality. Rather than occurring at a single critical point, the quadratic dispersion and critical nature of the material are protected over wide temperature ranges down towards absolute zero due to the special geometry of the crystal structure.

A second striking finding is the triad of magnetic phase transitions in $V_5S_8$ as a function of applied magnetic field below the Néel temperature. The field induced magnetic phase transitions point to the possible splitting of the spin-flop transition into two transitions, forming phases in which the alignment of the sublattice spins take on an unconventional form. The detailed nature of this spin alignment could be investigated by means of elastic scattering of polarized neutrons from the spin lattice in an applied magnetic field.

**Methods**

Single crystal specimens of $V_5S_8$ were grown by chemical vapour transport (CVT). In a typical procedure the reagents were prepared by mixing vanadium pieces with a purity of 99.99% and sulphur powder with a purity of 99.999% in the mass ratio appropriate for the desired stoichiometry. Vanadium pieces were pre-cleaned via acid etching and heating to high temperature in vacuum followed by a final acid-etching and cleaning step. Quartz growth ampoules were also pre-cleaned via acid etching followed by heating in vacuum to 1000°C for 24 hours to degas any impurities trapped inside the walls. The ampoules were etched again in a saturated solution of KOH in isopropyl alcohol followed by flame polishing the quartz. The reagents were then sealed together inside one end of an evacuated quartz ampoule, with an argon partial pressure, and iodine (purity 99.999%) as a transport agent at an amount of 20mg per cubic cm of the ampoule volume. The entire length of the sealed ampoule was heated to 675°C in a two-zone tubular furnace for a few hours. The unoccupied end of the ampoule, the growth end, was then cooled to 600°C to form a temperature gradient. Single-crystal growth was allowed to continue at the cold end for 15 days followed by slow cooling to room temperature over 2.5 days. Samples obtained were shiny hexagonal-shaped single-crystal platelets (see inset of Fig. 2e) of typical dimensions 1mm x 1mm x 0.2mm with the surfaces cleaned by acid etching in HCl before measurements of magnetic, electrical and thermal properties. The crystals were characterized by X-ray diffraction (XRD) and energy-dispersive X-ray spectroscopy (EDS) to confirm the crystal structure and stoichiometry (see also Appendix A). Susceptibility and magnetization measurements were performed using a commercial superconducting quantum interference device (SQUID) magnetometer with a setup to ensure alignment of the sample with the applied magnetic field. Resistivity and magneto-resistivity measurements were performed using a standard four-contact method at temperatures down to 0.5K and fields up 14T. Heat capacity measurements were carried out by a standard relaxation technique over the same temperature and field ranges.

## Acknowledgments

We would like to thank the Royal Society, the Henry Royce Institute under EPSRC grant number EP/R00661X/1, an EPSRC Doctoral Prize award (P. M. T. V.) and FAPESP under grant number 2017/10581-1 for financial support. SER would like to thank R. J. Cava and M. N. Ali for helpful discussions.

## Appendix A - X-ray diffraction (XRD) and energy-dispersive X-ray spectroscopy (EDS)

Single crystal samples of $V_5S_8$ were prepared by chemical vapour transport as described in the Methods section of the main text. The powder X-ray diffraction measurements confirmed that the single crystals were single phase and showed lattice parameters $a$ = 11.38Å, $b$ = 6.65Å, $c$ = 11.31Å, $\alpha$ = 90°, $\beta$ = 91.46°, $\gamma$ = 90° and volume = 855.24Å$^3$. The parameters match the space group F 1 2/m 1 having a monoclinic structure (also referred to as C1 2/m1 in modern nomenclature). The X-ray diffraction data were obtained from powdered material made from single crystals. After Rietveld refinement, it was possible to obtain an estimate of the stoichiometry of the atomic ratio of vanadium to sulphide of 5:7.95.

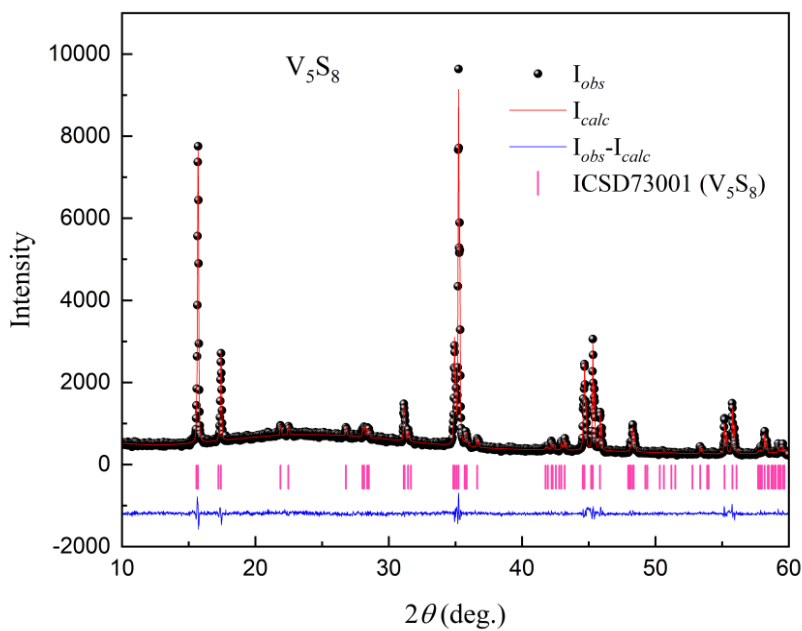

**Figure A1 - Powder diffractogram (black) and Rietveld refinement of the $V_5S_8$ phase (red). The positions of the observable peaks of the $V_5S_8$ phase are the magenta vertical bars. The blue line represents the difference between the observed pattern and the calculated one.**

Energy dispersive X-ray spectroscopy (EDS) of $V_5S_8$ was performed and gave an atomic ratio of vanadium to sulphur of 5:7.91 close to the result obtained from X-ray diffraction.